# DECAY ENERGY SPECTROMETRY FOR IMPROVED NUCLEAR MATERIAL ANALYSIS AT THE IAEA NML

G.B. Kim, A.R.L. Kavner, T. Parsons-Davis, S. Friedrich, O.B. Drury, D. Lee, X. Zhang, N. Hines
Lawrence Livermore National Laborat
Livermore, CA, USA
Email: kim90@llnl.gov

S.T.P. Boyd
University of New Mexico
Albuquerque, NM, USA

S. Weidenbenner, K. Schreiber, S. Martinson, C. Smith, D. McNeel, S. Salazar, K. Koehler, M. Carpenter, M. Croce
Los Alamos National Laboratory
Los Alamos, NM, USA

D. Schmidt, J. Ullom
National Institute of Standards and Technology
Boulder, CO, USA

**Abstract**

Decay energy spectrometry (DES) is a novel radiometric technique for high-precision analysis of nuclear materials. DES employs the unique thermal detection physics of cryogenic microcalorimeters with ultra-high energy resolution and 100% detection efficiency to accomplish high precision decay energy measurements. Low-activity nuclear samples of 1 Bq or less, and without chemical separation, are used to provide elemental and isotopic compositions in a single measurement. Isotopic ratio precisions of 1 ppm – 1,000 ppm (isotope dependent), which is close to that of the mass spectrometry, have been demonstrated in 12-hour DES measurements of ~5 Bq samples of certified reference materials of uranium (U) and plutonium (Pu). DES has very different systematic biases and uncertainties, as well as different sensitivities to nuclides, compared to mass-spectrometry techniques. Therefore, the accuracy and confidence of nuclear material assays can be improved by combining this new technique with existing mass-spectrometry techniques. Commercial-level DES techniques and equipment are being developed for the implementation of DES at the Nuclear Material Laboratory (NML) of International Atomic Energy Agency (IAEA) to provide complementary measurements to the existing technologies. The paper describes details of DES measurement methods, as well as DES precision and accuracy to U and Pu standard sources to discuss its capability in analysis of nuclear safeguards samples.

0. INTRODUCTION

The IAEA NML uses a variety of destructive and non-destructive techniques to detect diversion of nuclear materials [1]. The current baseline approach is isotope dilution mass spectrometry (IDMS) [2, 3]. In IDMS, a "spike" with well-known elemental and isotopic compositions is blended with the sample. The spike provides internal standardization and enables precise measurements of the blended sample to determine the elemental and isotopic composition of the original sample. While IDMS is a precise methodology, it has systematic uncertainties and biases, e. g. due to isobaric interferences, or due to uncertainty in the spike itself, as the compositions of the spikes are also determined by mass spectrometry techniques. Therefore, a separate and additional methodology with independent underlying uncertainties is very desirable to improve the overall accuracy of, and confidence in, the nuclear material analysis and material verification process for nuclear safeguards.

DES, also known as Q-spectrometry, is a novel radiometric technique that uses the unique thermal measurement physics of low-temperature microcalorimeters to directly measure the total energy change (Q) of a nucleus undergoing radioactive decay [4 – 6]. All radioactive actinides have distinct Q values and are easily distinguished by DES. Elemental and isotopic compositions can be simultaneously determined by a single DES measurement. DES uses very small samples with low activity of around 1 Bq. Typical precision achieved in a 12-hour measurement with a 1 Bq source is at the 1 ppm – 1,000 ppm level for the major isotopes in the nuclear sample, depending on the radionuclides present. This precision is close to that of IDMS, while the cost and time per sample for DES is significantly lower. Moreover, DES is a radiometric technique, with very different and independent systematic biases and uncertainties from those of mass-spectrometry. This makes DES a promising complementary measurement technology to IDMS, to improve the overall accuracy and confidence of measurements and thus to reduce uncertainties in nuclear verification. For this reason, the IAEA listed the development and deployment of DES to IAEA NML as one of the top priorities in the document "Development and Implementation Support Programme for Nuclear Verification 2022 – 2023" [7].

Lawrence Livermore National Laboratory (LLNL) and Los Alamos National Laboratory (LANL) are jointly developing technology for the routine DES analysis of actinide samples at the IAEA NML. Each laboratory is employing a different microcalorimeter technology: LLNL is developing DES using magnetic microcalorimeters (MMCs) [8, 9], while LANL is developing DES using transition edge sensors (TESs) [10, 11]. MMCs and TESs provide different strengths and systematic uncertainties for DES: MMCs provide faster counting speed, and TESs provide better energy resolution. Both technologies will be available in the DES instrument being jointly developed for deployment to the IAEA NML in 2025. Having two independent microcalorimeter technologies available will provide flexibility for measurement of different types of nuclear samples and help to establish confidence in DES, which is a new method for the IAEA NML.

1. METHOD

A DES detector consists of a gold foil that radioactive sources are embedded in and a cryogenic microcalorimeter to measure temperature changes of the gold foil by radioactive decays. An example DES setup is shown in Fig. 1. MMCs use gold wires and TESs use indium pads for the thermal contacts to the gold foil, but overall detection scheme is identical in both detectors.

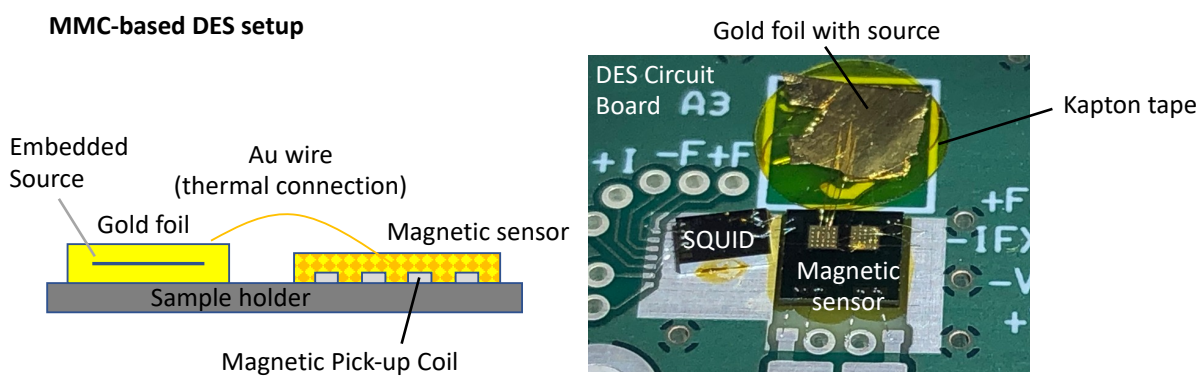

*Figure 1. (Left) A schematic drawing of a DES measurement with gold foil absorber and magnetic microcalorimeter (MMC). (Right) An example of a real MMC DES setup. The gold-foil absorber with source embedded is attached to a DES circuit board by Kapton tape. Gold wires are bonded between the gold-foil absorber and the MMC sensor to create the heat flow path enabling the read out of the decay energies. A quantum preamplifier ("SQUID," Superconducting QUantum Interference Device) amplifies the magnetic signals from the MMC sensor and convert it to voltage signals [12].*

A DES measurement consists of five steps: 1) sample preparation, 2) assembly with microcalorimeters, 3) cryocooling, 4) data acquisition and signal processing, and 5) spectral analysis to obtain material composition. These steps are described in the sub-sections below. A major focus of current DES development is standardizing, quantifying, and automating each DES measurement step to enable routine and reproducible DES measurements at the IAEA NML.

**Sample Preparation**

The first step of DES is embedding the actinide source into a gold foil absorber. Proper embedding is essential for best performance in DES. The radioisotope sample needs to be finely divided and well-mixed with the gold foil matrix to ensure that, as much as possible, decay products deposit their energy directly into the gold. Any fraction of the decay energy that is instead self-absorbed by the sample material can have a long time constant for appearing in the thermal signals, and this degrades energy resolution and increases systematic uncertainties. This self-absorption concern is similar to that of alpha spectroscopy. A second concern for sample preparation is that cross-contamination of samples might occur if the embedding is not carefully performed. For these reasons, the development of reproducible and reliable sample-embedding technique is a significant focus of current DES technology development.

Importance of the sample embedding technique can be quantified by considering the three major factors determining DES energy resolution, expressed as

$$\sigma_{\text{tot}}^2 = \sigma_{\text{det}}^2 + \sigma_{\text{source}}^2 + \sigma_{\text{drift}}^2, \qquad (1)$$

where $\sigma_{\text{tot}}$ is the total DES energy resolution, $\sigma_{\text{det}}$ is the detector's intrinsic energy resolution, $\sigma_{\text{source}}$ is the broadening due to source self-absorption, and $\sigma_{\text{drift}}$ is the broadening effect caused by the detector's gain drift due to fluctuation of the operating temperature. The detector's intrinsic energy resolution $\sigma_{\text{det}}$ is set by the total heat capacity of the DES detector, which is mainly determined by the volume of the gold foil. This term is approximately 30 eV at 50 mK operating temperature for both MMCs and TESs, if a gold foil of 3 mm × 3 mm × 25 μm is used [13]. This is significantly better than typical DES resolutions of a few keV, and therefore $\sigma_{\text{det}}$ is not a dominant factor determining -DES resolution unless significantly larger gold foils are used. If it is not compensated for, the gain-drift term $\sigma_{\text{drift}}$ can cause significant broadening of tens of keV. However effective correction algorithms have been developed (see "Data acquisition and signal processing" below), reducing the impact of $\sigma_{\text{drift}}$ on energy resolution well below $\sigma_{\text{source}}$. For poorly-embedded sources, $\sigma_{\text{source}}$ can impact energy-resolution at the 1 keV-10 keV level [6].

Various techniques have been developed to improve $\sigma_{\text{source}}$ by breaking up the source microcrystals. In the baseline approach, the actinide source is dissolved in acid solution, then pipped onto a gold foil. The acid is evaporated, leaving only the source on the foil. After complete evaporation, the gold foil is folded over so that the source is fully enclosed. The gold foil is then "kneaded," *i.e.*, folded, reoriented, and pressed multiple times with a smooth-jawed pliers and isolating titanium foil, to break down the source microcrystals. With 100 kneading cycles, this technique has demonstrated the highest DES energy resolution, yielding 1.0 keV full-width-half-maximum for the $^{239}$Pu peak [6].

In addition to the baseline technique, we are actively developing two additional approaches for quick, reliable, and reproducible sample preparation at nuclear safeguards laboratories. The first approach, being developed at LLNL, is to perform the mechanical alloying with a small hand-operated rolling mill. This approach is otherwise similar to the high-performing baseline approach but replaces the kneading process with a quick and reliable means of mechanical alloying. The second approach, being developed at LANL, uses nano-porous gold foils. The acid solution soaks into the small pores of these metal foils, and the pore size sets an upper bound on the size of the source material crystals.

Figure 2 summarizes the rolling mill procedure and the result of a single rolling cycle. The folded gold/source/gold is sandwiched between two stainless steel (SS) foils to protect the gold foil and the rolling mill from cross-contamination. This sandwich is then inserted into a small hand-operated rolling mill. The folded gold/source/gold package is uniformly thinned and lengthened in the rolling mill. Package thickness is adjusted by changing the roller separation distance. The stretching of the gold breaks up any crystals in the source material while simultaneously embedding each source particle intimately into the gold. Further, the two layers of gold are fully cold-welded together, forming a sealed absorber for further ease of handling. This procedure can be repeated by folding the rolled-out gold foils and re-rolling them. There are three advantages of this rolling mill method: 1) speed – the entire procedure takes less than 5 minutes; 2) the starting and rolled-out thicknesses of the gold foil are easily controlled for quality control and reproducibility, and 3) any possible cross-contamination between different samples can be eliminated by discarding the used stainless-steel foils.

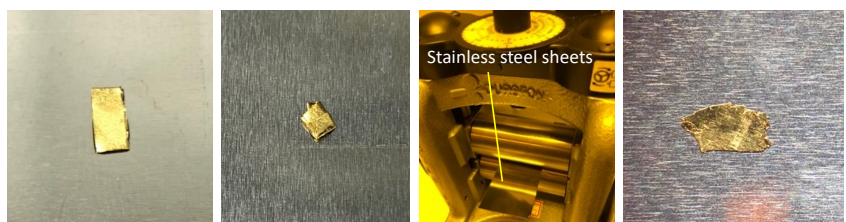

*Figure 2. Mechanical alloying method using a hand-operated rolling mill. From left to right, photographs show the gold foil after source deposition, after folding, the rolling mill with the gold foil sample sandwiched by stainless steel sheets for preventing contamination, and the rolled-out gold foil as a result.*

The technique using nano-porous gold foil requires aqueous solution samples to be pipetted onto nano-porous gold layers which constrain the size of crystals that form as the solution dries. This creates a nanocomposite material with optimized thermal performance. In contrast to the baseline "kneading" technique of mechanical alloying, solution deposition into nano-porous gold would be extremely simple to implement.

Figure 3 *(Left)* shows a scanning electron microscope image of the nano-porous gold. This material is made by co-sputtering Au and Ag in approximately a 1:1 ratio on top of a standard Au foil, then removing the Ag with nitric acid to leave a nano-porous Au film. Pieces can be easily cut and attached to a microcalorimeter sensor. The typical pore size of 50-100 nm provides adequate homogeneity of the sample to resolve $^{240}$Pu/$^{239}$Pu in the decay energy spectrum (Figure 3 *(Right)*). It is desirable to further reduce low-energy tails in the spectrum, which may come from energy loss in the sample material due to the relatively large pore size.

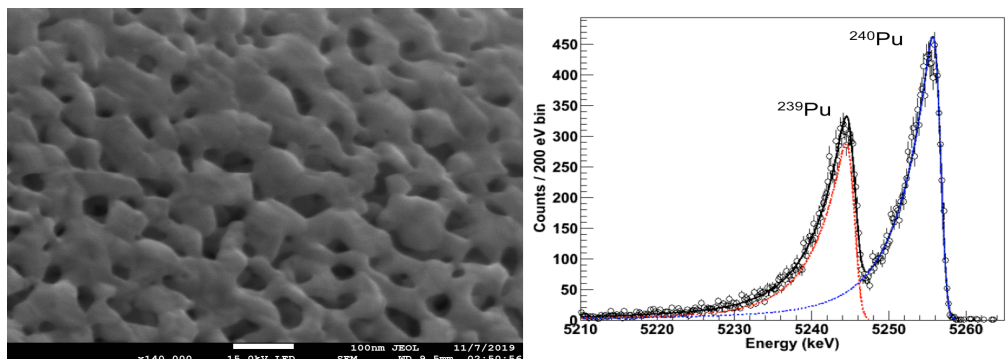

*Figure 3: Nano-porous gold material and DES spectrum. (Left) SEM image with 100nm scale bar. Nano-porous gold with 50-100 nm pore size provides adequate energy resolution and is easy to handle due to its Au foil backing. (Right) Decay energy spectrum resulting from a solution of Pu in nitric acid pipetted onto the nanoporous material.*

In addition to breaking up and embedding the source microcrystals, there are other considerations in sample preparation. If mechanical alloying cycles are performed too many times, source material might be exposed on the surfaces of the gold foils or embedded only under layers of gold ≲ 10 um thick, and this could lead to alpha-escapes and energy losses. Such energy losses would produce low energy tails or flat backgrounds, creating systematic biases in spectral analysis. Surface source exposure could potentially also cause cross-contamination between different samples, or contamination of the microcalorimeters or the mounting holders, leading to background signals in subsequent measurements. As DES uses sources with only ~1 Bq total activity, such background signals may not be immediately apparent. However, they could still cause biases on analysis results. To avoid surface contamination and energy loss issues, both the number of mechanical alloying cycles and thickness of gold foils are under investigation.

A further consideration is the ability to embed proportionally large amounts of source into the absorber, and this problem applies across both the mechanical alloying and nano-porous gold approaches. To achieve the best energy resolution, the mass of gold foil or nano-porous gold should be minimized. However, because of U isotopes' long half-lives, as much as 50 μg of U source might be required for adequate activity and statistics in a 12-hour DES measurement. That is approximately 1% of the mass of a gold foil of 3mm × 3mm × 25μm. It may prove difficult to fully embed the source for that large mass ratio using mechanical alloying, or to load that much solution into the nano-porous gold. Such loading may significantly change the thermal properties of the gold foil or nano-porous gold and degrade detector performance. If embedding or loading these large amounts of material for the case of U does prove to be a problem, two solutions immediately suggest themselves: 1) use smaller activity but longer measurement times, or 2) use larger absorbers at lower resolution. DES spectral analysis research at LLNL suggests that in many cases count rate is more important than energy resolution for determining isotope ratios as long as peaks are separated (see "Result and DES performance" section below), so case 2) may be preferred. Lastly, we note that this is only a problem for U-bearing samples. Pu's shorter half-lives require much smaller masses for adequate activity, and do not create this problem.

For all these reasons, optimal parameters and procedures for DES sample preparation are being determined by carefully considering and measuring multiple factors: heat capacities and detector energy resolution, total mass of the source, potential for cross-contamination, and the ease and reproducibility of mechanical alloying processes or solution loading into nano-porous gold.

**Microcalorimeters and Detector Assembly**

To precisely measure the thermal signals created by decay energy absorption, the gold foil absorbers containing their embedded samples must then be mounted to cryogenic microcalorimeters. Both MMCs and TESs are operated at very low temperatures, typically less than 0.1K. MMCs use paramagnetic sensor materials with their $M = 1/T$ Curie-Law response (where *M* is magnetization and *T* is temperature) to measure the temperature increase from the decay energy, while TESs use a superconducting film and measure resistance changes near the superconducting transition temperature. In both cases, a small temperature increase ($\Delta T$) results in a large sensor response. Sensitivities of these microcalorimeters are better than 0.1 μK, yielding resolutions typically between 1/1000 to 1/5000. This is sufficient to fully resolve most of the actinide peaks in decay energy spectra. MMCs are relatively faster and more linear, thus are appropriate for high accuracy measurements with high statistics. TESs generally exhibit better energy resolution in DES applications and thus are appropriate for measurements requiring high resolving power.

For MMC microcalorimeters, the gold foil absorbers are first attached to a circuit board with 2-sided Kapton tape, and then coupled to the MMC via gold wire bonds for a fast thermal connection (Fig. 4). After a measurement, the Kapton tape can be removed together with the absorber to prevent cross-contamination. For TES microcalorimeters, the gold foil absorbers are attached with a small piece of indium to a micromachined silicon chip containing the TES. There will be four to eight detector pixels per circuit board. The circuit board is then mounted into magnetic shielding to reduce noise levels, and the resulting package is mounted to the cooling platform. An example MMC DES circuit board and mounting scheme is shown in Fig. 4.

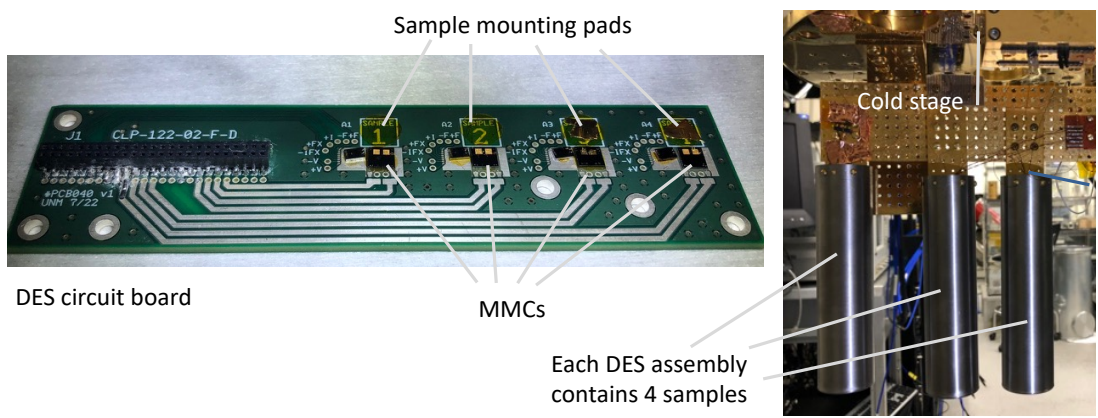

*Figure 4. (Left) DES circuit board with MMCs and gold foil samples. (Right) The magnetically-shielded detector packages are installed onto the cold stage of a dilution refrigerator.*

For the DES instrument at the IAEA NML, 16 detector pixels consisting of 8 MMCs and 8 TESs will be mounted in a single instrument so that 16 nuclear samples can be simultaneously measured in one measurement. Both types of microcalorimeters use similar quantum preamplifiers, the SQUIDs, and similar room-temperature readout electronics. SQUIDs and their readout electronics are commercially available and are well-proven technologies. Their use should be essentially "transparent" to the end user.

**Cryocooling**

Both the MMC and TES approaches to DES require cryogenic operating temperatures, $T < 0.1$ K. There are two types of commercially available cryostats that can reach this temperature range: adiabatic demagnetization refrigerators (ADRs [14]), and dilution refrigerators (DRs [15]).

ADRs rely on the relatively large changes of entropy and temperature that can be induced in paramagnetic materials at low temperature by applying or removing a magnetic field. Alternating the high and low magnetic field on a paramagnetic refrigerant is sequenced with the opening and closing of thermal switches to create sub-K refrigeration. ADRs generally have a compact footprint and can be operated without special electric power or water cooling and are therefore appropriate for laboratories looking for a compact system for DES. An advantage of ADRs is that the sample stage is decoupled from the higher-temperature stages of the cryostat both mechanically and thermally, providing very quiet conditions for the DES measurement. A disadvantage of ADR is the cyclic, rather than continuous, nature of the cooling: the needed DES operating temperature can be achieved for less than 24 hours per cycle. Longer measurements must be conducted through several cooling cycles.

Dilution refrigerators (DRs) rely on phase separation phenomena in mixtures of superfluid $^3$He and $^4$He. They provide continuous cooling power at greater levels than a typical ADR. However, DRs are larger than ADRs, require large amounts of increasingly scarce and expensive $^3$He to operate, and have greater infrastructure requirements for power and cooling water.

**Data acquisition and signal processing**

We are developing a commercial-grade, user-friendly automatic data acquisition (DAQ) system for the new DES instrument. The DES signals, from both MMCs and TESs, are read out by SQUID preamplifiers and electronics. Biasing and tuning of the SQUIDs are automated and do not require user inputs. MMCs and TESs also require persistent current injection and biasing currents, respectively, which are implemented by automated processes in the control software. DES signal pulses can be triggered using common trigger algorithms such as a rising-edge trigger. Captured pulses are processed by two types of filter algorithms to extract pulse amplitudes. MMC signals are processed by fast trapezoidal filters that are commonly employed in semiconductor detector systems [16], and TES signals are processed by optimal filters [17] that provide the best energy resolution. Signals are currently processed by off-line algorithms. However, online signal processing software is in development for the commercial DES instrument. Triggering, filtering, and amplitude calculation will be automatically performed with minimum user input by online data acquisition and signal processing software, giving functionality similar to the multi-channel analysers (MCAs) of semiconductor detector instruments. DES energy spectra will be obtained in real time during data acquisition.

Compared to semiconductor detectors, DES requires an additional data-analysis step to correct for amplitude drift caused by temperature fluctuations. However, effective correction algorithms have been developed to compensate for these drifts. These algorithms and DAQ hardware are well-proven in research settings. As shown in Figure 5, a moving-median is calculated for the strongest peak, and the gain is normalized to the moving-median trace. This drift correction can be performed by an automatic algorithm after the data acquisition is completed, allowing further improvement of DES performance.

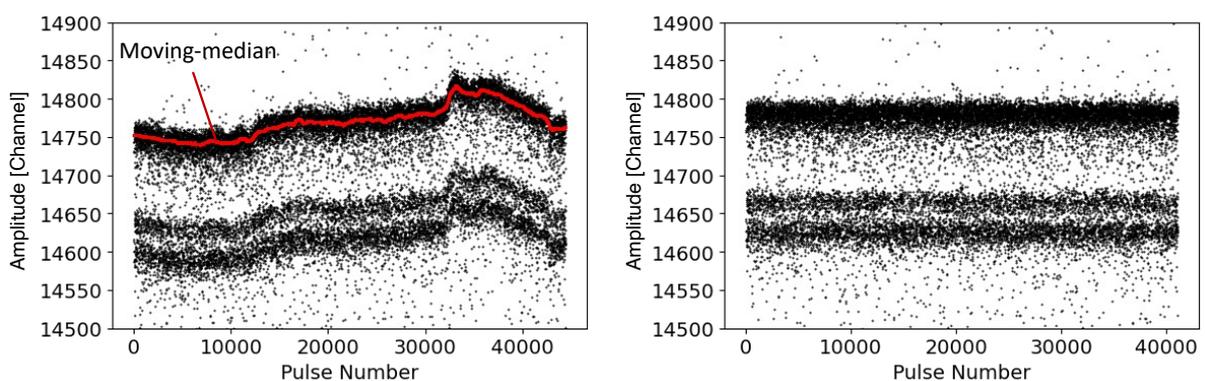

*Figure 5. (Left) A scatter plot showing the drift of signal amplitudes. The calculated moving-median along a strong peak is shown in red. (Right) After applying the median-filter correction.*

**Spectral Analysis**

After pulse processing and drift correction, the measured signal amplitudes are calibrated to known peak energies, and energy spectra are obtained from histograms of amplitude distributions. Example DES spectra of U and Pu sources are shown in Fig. 6. Analyzing DES spectra is relatively straightforward compared to those from other radiation detectors, because there is very little background, the high energy resolution gives minimal line overlap, and most of the actinide isotopes produce single peaks rather than the multiple peaks seen with alpha spectrometers. The elemental and isotopic ratios are extracted from the number of counts in each decay energy peak, taking half-lives into account. Thus, extracting peak counts is the most important work in DES spectral analysis. Peak counts are obtained by spectral fitting. Therefore, understanding and using the right fitting functions is essential for minimizing systematic uncertainties and achieving high accuracy isotopic measurements. DES peak shapes are well-described by the sum of two exponentially modified gaussian functions [18], the convolution of Gaussian and exponential functions,

$$f(E) = EMG(E; a_1, E_0, \sigma, P_1) + EMG(E; a_2, E_0, \sigma, P_2), \text{ and} \quad (2)$$

$$EMG(E; a, E_0, \sigma, P) = \frac{a\sigma}{P}\sqrt{\frac{\pi}{2}} \exp\left(\frac{1}{2}\left(\frac{\sigma}{P}\right)^2 - \frac{E-E_0}{P}\right) \text{erfc}\left(\frac{1}{\sqrt{2}}\left(\frac{\sigma}{P} - \frac{E-E_0}{\sigma}\right)\right), \quad (3)$$

where "erfc" is the complementary error function, $E$ is energy, $a$ is amplitude, $E_0$ is the peak centroid, $\sigma$ is the standard deviation of the original gaussian function, and $P$ is the tailing parameter created by the exponential function. Parameters $a_1$ and $a_2$ are scaled so that the area of $f(E)$ is normalized to 1. A third tail component may be added if needed, especially in U decay energy spectra with larger tails. All DES peaks can be fitted by summing a series of $f(E)$, e.g.,

$$F(E) = \sum A_i f_i(E; a_1, a_2, E_{0i}, \sigma, P_1, P_2). \quad (4)$$

where $F(E)$ is the total fit function of the entire DES spectrum and $A_i$ is the area or total counts of each decay energy peak $i$ in the DES spectrum. All decay energy peaks share the same $a$, c, and $P$ values, as well as $a_1$ to $a_2$ ratios, therefore only the $A_i$ and $E_0$, the peak areas and centroids for each decay energy peak, are independent parameters. Total peak counts, or the total number of decays of each isotope, are obtained by integrating the fit function area. The fit residuals should be consistent with zero, otherwise integrated peak counts can differ from true peak counts and produce systematic biases in isotopic ratio analysis. Consistency of peak shapes between data and fitting function, and between different decay energy peaks at different energies, is one of the most important factors determining systematic uncertainties in DES analysis.

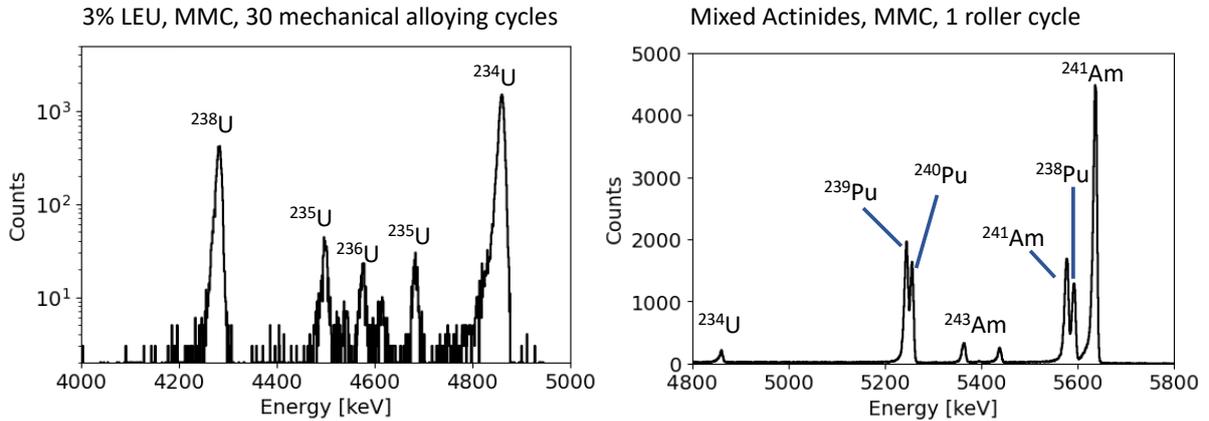

*Figure 6. (Left) Measured DES spectrum of 3% low enriched uranium (LEU) source measured by MMCs. 30 mechanical alloying ("kneading") cycles were used to break up the source microcrystals. (Right) MMC DES spectrum of a mixed actinide source with a single rolling-mill cycle. Here the mechanical alloying was accomplished with a single pass through the roller. Multiple elements are identified in a single measurement.*

Analyzing Pu decay energy spectra is relatively straightforward as most Pu isotopes produce single peaks. Am isotopes produce "escape peaks" due to their strong gamma-ray emissions, *e.g.* 60 keV gamma-rays from $^{241}$Am. In such cases the primary and escape peaks should be summed to obtain the total decay counts of $^{241}$Am. The analysis of U decay spectra is more involved. $^{235}$U has a large number of escape peaks and produces a continuum over a wide energy range of a few hundred keV.

LLNL has developed a novel fitting method for the accurate analysis of U decay energy spectra. Nuclear decays of U and Pu, and radiation transport in the gold foil are simulated by Geant4 for various gold foil thicknesses and source distributions. Energy deposition to the gold foil is collected for one million simulated decays of each U and Pu isotope to reproduce the DES detector's response. With this new method, escape peak structures of $^{235}$U and other isotopes emerge naturally, as shown in Fig. 7. The simulated spectra are convolved with a flexible peak shape function (Eq. 2), and those convolved spectra are then fitted to the DES spectrum to extract U isotopic composition. The fitting function can be mathematically expressed as

$$f'(E) = \int f(E; a_1, a_2, E_{0i}, \sigma, P_1, P_2) G_i(E') dE', \text{ and} \tag{5}$$

$$F'(E) = \sum A_i f_i'(E; a_1, a_2, E_{0i}, \sigma, P_1, P_2), \tag{6}$$

where $f(E)$ is the peak shape function in Eq. 2, $G_i(E')$ is the simulated spectral distribution of the i-th isotope. Then the fitting function $F'(E)$ is used to fit the experimental decay energy spectrum to find the best values for $A_i$, $E_{0i}$, and the peak shape parameters. As all peaks share the same shape, ratios and uncertainties of $A_i$ can be used for extracting the isotopic ratios. An example decay energy spectrum and fit result using this method are shown in Fig. 8. This new method represented the $^{235}$U escape peak structure well and the fitted result is consistent with the data, whereas the standard fitting method of individual peaks cannot count all $^{235}$U decay events. This method does not require online Geant4 simulation. A set of simulated spectral data will be embedded in the fitting algorithm and used as internal library data, thus Geant4 operation is not needed for routine DES analysis.

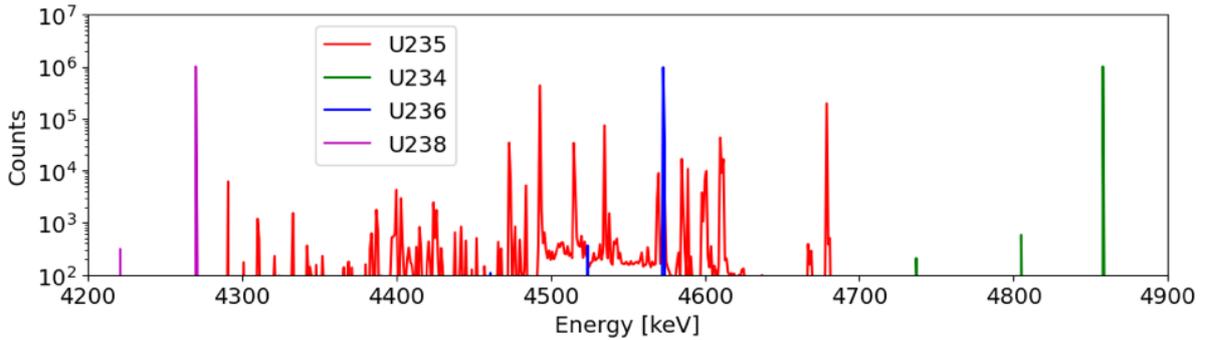

Figure 7. Simulated decay energy spectra of U isotopes. Minor lines are due to gamma-ray escapes from the gold foil. $^{235}$U exhibits a large number of escape peaks due to its high intensity gamma-rays, whereas $^{234}$U, $^{236}$U, and $^{238}$U are mostly forming single peaks at their total decay energy (Q) values. The $^{235}$U spectral structure should be taken into account for accurate isotopic composition analysis.

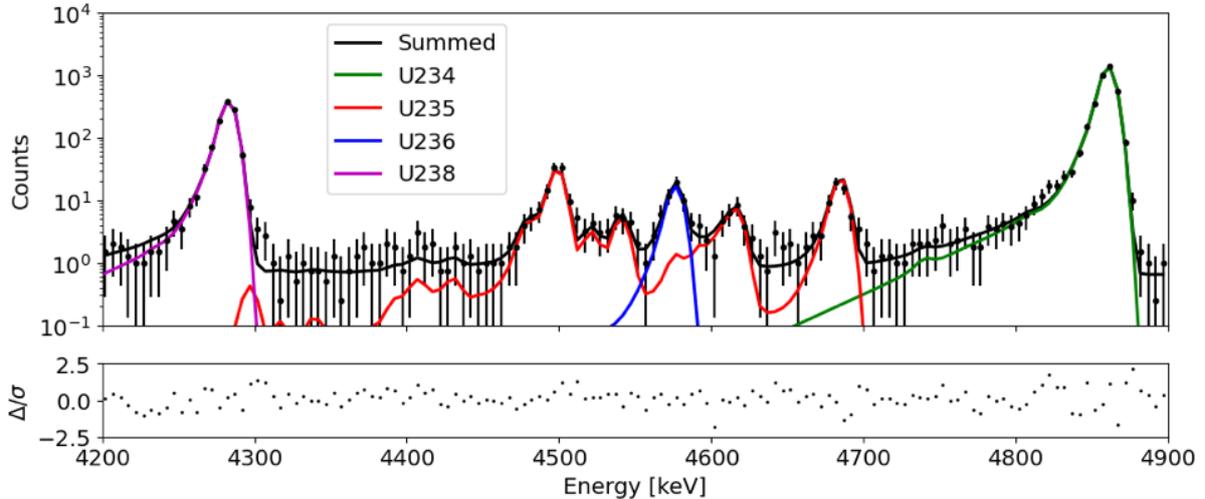

Figure 8. Decay energy spectrum of a 3% low-enriched U source and spectral fitting result using the Geant4 simulation technique to include escape peak structures of $^{235}$U. Δ/σ denotes fit residuals in unit of standard deviation. Data and 235U escape peak structure are represented well by the fitting function from the new method using Geant4 simulation.

## 2. RESULT AND DES PERFORMANCE

The key performance metric for the use of DES at nuclear materials laboratoriesNML is the accuracy and precision of isotopic composition measurement. This has been tested at LLNL with MMCs for the U and Pu standard sources CRM137A, U100, and U030 [19]. Multiple measurements were conducted for each source type to investigate measurement accuracy and systematic uncertainties. The activity of sources was also varied to investigate any accuracy dependence on the source amount of gold foil volumes. Gold foils were treated by the mechanical alloying method and measurements were performed in a DR. MMC signals were processed using a trapezoidal filter with the drift correction described above.

Figure 9 shows measurement precision versus measurement times for MMC-based DES for various U and Pu isotopes in the standard sources. The statistical precision of the $^{235}$U/$^{238}$U ratio reached $10^{-3}$ level within 10 hours of DES measurement. The precisions of the $^{234}$U/$^{238}$U and $^{236}$U/$^{238}$U ratios were significantly better than that of $^{235}$U/$^{238}$U, due to the shorter half-lives and higher activity of $^{234}$U and $^{236}$U. This is an important characteristic of the DES technique: DES is a radiometric technique and its sensitivities to isotopes are different from those of mass spectrometry. Thus, DES is a promising complementary technology for nuclear material analysis.

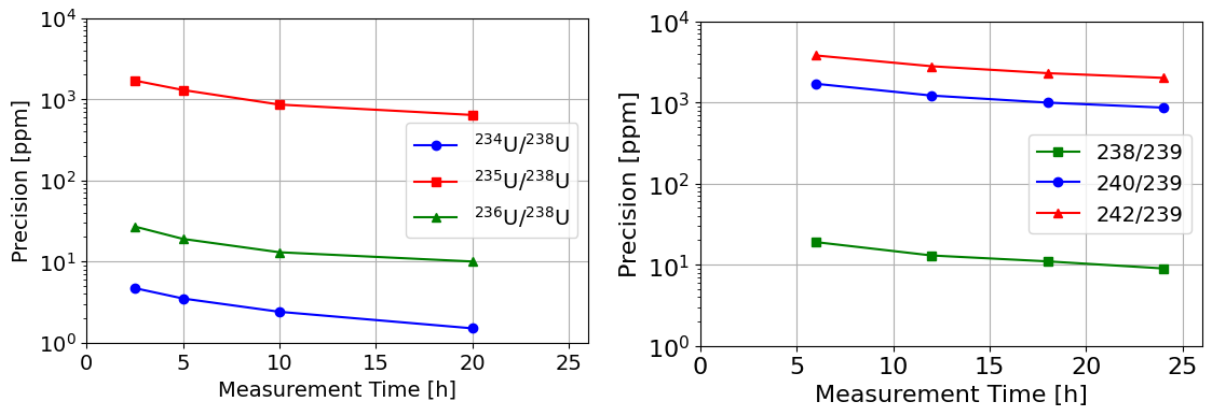

*Figure 9. Measured DES isotope ratio precision versus measurement time. (Left) For U030 U, using a 6 Bq source. (Right) For Pu (CRM137A), using a 4 Bq source.*

Measurement accuracy has been investigated by comparing a series of measurements using different sources, gold foil geometries, source activities, and different MMC sensors having different sensitivities. Table 1 lists the measured isotopic ratios for U030. The U030 reference values are shown in the top row for comparison. Most of DES analysis results agree with the certificate numbers within statistical uncertainties, whereas $^{234}$U/$^{238}$U were not, which indicate that there are systematic uncertainties of an order of 0.001%. The DES measurement time can be adjusted to achieve the required precision for each nuclear sample.

TABLE 1. MMC-DES analysis results for various samples using U100 and U030 reference materials. Numbers are in atomic ratios. Uncertainties for DES measurements are statistical. Half-life values are employed from Nuclear Data Sheets (NDS) [20 – 23].

|  | Sample Prep. | $^{234}$U/$^{238}$U [%] | $^{235}$U/$^{238}$U [%] | $^{236}$U/$^{238}$U [%] |
|---|---|---|---|---|
| U100 |  | 0.0753(2) | 11.36(1) | 0.0422(1) |
| Run46 | 1 Folding | 0.0810(8) | 11.22(13) | 0.046(4) |
| Run47 | 12 Folding | 0.0774(8) | 11.54(19) | 0.044(2) |
| Run48 | 12 Folding | 0.0759(5) | 11.38(11) | 0.043(1) |
|  |  |  |  |  |
| U030 |  | 0.0196(1) | 3.147(3) | 0.0206(1) |
| Run54 | 30 Folding | 0.0190(11) | 3.07(18) | 0.0190(11) |
| Run56 | 3 Folding | 0.0196(2) | 3.02(5) | 0.0206(10) |
| Run58 | 10 Folding | 0.0207(2) | 3.22(9) | 0.0196(10) |

DES performance for Pu analysis was also experimentally examined with the MMC using a CRM137A (freshly separated CRM137) source of 4 Bq and 0.6 Bq activities, respectively. The 0.6 Bq source was prepared with 10 folding cycles whereas the 4 Bq source was prepared with only 3 folding cycles to test DES performance and accuracy when sample preparation is minimal and when a high-activity source is used. The 0.6 Bq sample therefore exhibited a better energy resolution. Most of measured isotopic ratios were consistent with the certificate composition of the CRM137 in both cases within statistical uncertainties. Statistical precision was higher for the 4 Bq sample due to its higher statistics. The second row of Tab. 1 is calculated using the CRM137 composition certificate with age correction. It is noted that DES precisions for the $^{238}$Pu/$^{239}$Pu ratio were better than that of the certificate.

The Pu measurements show another advantage of using DES. The source contains not only alpha-decaying actinides but also beta-decaying $^{241}$Pu. Both low energy beta decay signals from $^{241}$Pu and MeV-scale alpha decay signals from other Pu and Am isotopes were obtained in a single measurement. DES provides an un-biased analysis for inter-elemental ratios, thus the $^{241}$Pu/$^{241}$Am ratio can be obtained directly for Pu-Am chronometry. Total counts of $^{241}$Pu beta decays were obtained by fitting a simulated theoretical curve to the $^{241}$Pu beta spectrum experimental data. Using the $^{241}$Am/$^{241}$Pu ratio allows for the sample separation date to be estimated using a nuclear dating technique [24]. The estimated separation date by DES was Sep-21 2021 ± 20 days, which was consistent with the true separation date of this sample, Sep-28 2021.

TABLE 2. Calculated isotopic ratios from CRM137A and DES measurement results for 24 hours for samples of two different activities and energy resolutions with different sample preparation. Measured precision for the $^{238}$Pu/$^{239}$Pu ratio is better than that of the certificate. The calculated $^{241}$Am/$^{239}$Pu ratio is not listed as it is different for two measurements that are done in different days. $^{241}$Pu has 20% systematic uncertainty due to event cut efficiency. Other uncertainties are all statistical. Half-life values are employed from NDS [25 - 30].

|        | Sample Prep. | ΔE    | $^{238}$Pu/$^{239}$Pu [%] | $^{240}$Pu/$^{239}$Pu [%] | $^{241}$Pu/$^{239}$Pu [%] | $^{242}$Pu/$^{239}$Pu [%] | $^{241}$Am/$^{239}$Pu [%] |
|--------|--------------|-------|---------|---------|-----------|---------|-----------|
| CRM137 |              |       | 0.264(6) | 24.07(3) | 0.5258(15) | 1.561(5) |           |
| 4 Bq   | 3 Folding    | 8 keV | 0.2604(14) | 23.28(13) | 0.49(10) | 1.54(13) | 0.0049(4) |
| 0.6Bq  | 10 Folding   | 5 keV | 0.266(2) | 24.0(2) | 0.52(10) | 1.7(3) | 0.0050(5) |

3. CONCLUSION

DES provides high precision and accuracy measurements using low mass, low activity samples without the need of chemical separation. It complements current safeguards sample analysis techniques, as it provides isotope analysis with very different systematic biases, uncertainties, and sensitivities. The DES operating procedures have been standardized and simplified for routine measurements at IAEA NML. Potential systematic biases, uncertainties, and cross-contamination factors are carefully examined for controlled, reproducible, and straightforward sample preparation procedures. Data acquisition and spectral analysis software are being developed for user-friendly operation and analysis.

To verify DES precisions and accuracies, a series of measurements were conducted with various standard U and Pu sources. Measurement results from different sample preparation methods and number of cycles, activities, detector resolutions, and counting statistics are compared. DES provided consistent results for isotopic ratios in all measurements within statistical uncertainties that are varying from 1 ppm to 1,000 ppm for different isotopes. This shows that DES can provide high accuracy analysis for various actinide samples within 24 hours of measurement time. In addition, thanks to the DES's unbiased measurement capability for different elements in a single measurement, $^{241}$Pu/$^{241}$Am chronometry is available to estimate sample ages.

**ACKNOWLEDGEMENTS**

(Contract No. 89233218CNA000001). The work presented in this paper was funded by the National Nuclear Security Administration of the Department of Energy, Office of International Nuclear Safeguards.